\documentclass[useAMS,usenatbib]{mn2e}
\usepackage{times}
\usepackage{graphics}
\def\etal{{et al.}}

\def\eff{{e\!f\!f}}

\def\gtwodeff{g_{2D\!,e\!\;\!f\!f}}

\title[Genus Topology of the Cosmic Microwave Background from the WMAP
  3-Year Data]{Genus Topology of the Cosmic Microwave Background from
  the WMAP 3-Year Data}
\author[J. Richard Gott, III,
        Wesley N. Colley,
        Chan-Gyung Park,
        Changbom Park and
        Charles Mugnolo]
{J. Richard Gott, III$^1$,
 Wesley N. Colley$^2$\thanks{E-mail: colleyw@uah.edu},
 Chan-Gyung Park$^3$, Changbom Park$^3$, and
\newauthor Charles Mugnolo$^1$ \\
$^1$Dept. of Astrophysical Sciences, Princeton University,
    Peyton Hall, Ivy Lane, Princeton, NJ 08544\\
$^2$Center for Modeling, Simulation and Analysis, D-15 Von Braun
    Research Hall, Univ. of Alabama in Huntsville, Huntsville, AL 35899 \\
$^3$Korea Institute for Advanced Study, 207-43 Cheong-Yang-Ni,
    Dong-Dae-Mun, Seoul 130-722, Korea
}

\begin{document}
\date{Accepted . Received }

\pagerange{\pageref{firstpage}--\pageref{lastpage}} \pubyear{2006}

\maketitle

\label{firstpage}

\begin{abstract}
We have independently measured the genus topology of the temperature
fluctuations in the cosmic microwave background seen in the Wilkinson
Microwave Anisotropy Probe (WMAP) 3-year data.  A genus analysis of
the WMAP data indicates consistency with Gaussian random-phase initial
conditions, as predicted by standard inflation.  We set 95\%
confidence limits on non-linearities of $-101 < f_{nl} < 107$.  We
also find that the observed low $\ell$ ($\ell \le 8$) modes show a
slight anti-correlation with the Galactic foreground, but not
exceeding 95\% confidence, and that the topology defined by these
modes is consistent with that of a Gaussian random-phase distribution
(within 95\% confidence).
\end{abstract}


\section{Introduction}

The Wilkinson Anisotropy Probe (WMAP) has revolutionized our
understanding of the cosmic microwave background (CMB) anisotropy.  In
2003, the team produced its first data release (Bennett~\etal\ 2003a)
based on one year of observations.  The latest release (Hinshaw~\etal\
2006) features almost 3 years worth of observations, dramatically
increasing the signal-to-noise of the CMB maps.

As with the previous data release, the team has formed best estimates
of cosmological parameters based on the data in various combinations
with other cosmological information (Spergel~\etal\ 2006), such as
Type Ia supernovae (e.g., Riess~\etal~2004), elemental abundances
predicted by Big Bang nucleosynthesis, and the Hubble Key Project
(Freedman~\etal\ 2001), among many others.  The tool of choice for
assessing these cosmological parameters is the power spectrum of the
observed fluctuations in the CMB sky.  This is computed by calculating
the products of the spherical harmonic coefficients $a_{\ell m}$ and
their complex conjugates; then, for each $\ell$, those products are
summed over all $m$ values to give the total power, $C_\ell$, at that
angular scale.  Different cosmologies predict quite different
power-spectra, and so the real data can be tested against predicted
power-spectra to give high-confidence estimates of the cosmological
parameters.

The $a_{\ell m}a_{\ell m}^*$ product in the power spectrum, however,
explicitly removes phase information in the spherical harmonic modes.
These phases contain critical information for characterizing the
primordial density fluctuations.  Namely, standard inflation (e.g.,
Guth 1981; Albrecht \& Steinhardt 1982; Linde 1982, Linde 1983)
predicts that the temperature fluctuations in the CMB, at the
resolution measured by WMAP, will be characterized by spherical
harmonic coefficients with Gaussian-distributed amplitudes and random
complex phases.  The WMAP data provide our best opportunity to date to
test that hypothesis.

The genus topology method developed by Gott, Melott \& Dickinson
(1986) directly tests for the Gaussian random-phase nature of a
density (or temperature) distribution in 3 dimensions (Adler 1981;
Gott, Melott \& Dickinson 1986; Hamilton, Gott \& Weinberg 1986; Gott,
Weinberg \& Melott 1987), or in 2 dimensions (Adler 1981; Melott
\etal\ 1989). Coles (1988) independently developed an equivalent
statistic in 2 dimensions.  The 2 dimensional case has been studied
for a variety of cosmological datasets: on redshift slices (Park
\etal\ 1992; Colley 1997; Hoyle, Vogeley \& Gott 2002), on sky maps
(Gott \etal\ 1992; Park, Gott, \& Choi 2001), and on the CMB, in
particular (Gott et al. 1990; Smoot \etal\ 1992; Kogut 1993; Kogut
\etal\ 1996; Colley, Gott \& Park 1996; Park \etal\ 1998; Park,
C-G. \etal\ 2001).  Watts \& Coles (2003) and Chiang \& Coles (2000),
among many others, have investigated other methods for measuring
phases, such as looking at the Fourier modes directly.

The WMAP team has carefully measured the genus of the WMAP sky as seen
in the new 3-year data (Spergel~\etal\ 2006), and demonstrated that
the WMAP results are consistent with the Gaussian random-phase
hypothesis.  To do this, they carried out a large number of
simulations of the CMB, in which the spherical harmonic coefficients
were drawn from a Gaussian random-phase distribution.  They then used
their known beam profiles to synthesize the results in each frequency,
and applied the {\it Kp0} (Hinshaw~\etal\ 2006) mask, just as one
would with the real dataset.  Using statistical techniques to be
discussed at length herein, they determined that the WMAP data are
consistent with a Gaussian random-phase field, as predicted by
Inflation.  They carried out further tests, using other Minkowski
functionals (Minkowski 1903; the genus is one functional, others are
the area fraction and the contour length) and the bispectrum to verify
the Gaussian random-phase hypothesis.

We seek to confirm this result using our own methods.  Rather than
comparing to simulated Gaussian random-phase realizations of the CMB,
we compare data directly to the theoretical prediction for the
random-phase genus in two dimensions.

\section{WMAP Observations: Maps}

First, we will plot the WMAP data using a color scheme developed by
Colley \& Gott (2003) and using some new map projections developed by
Gott, Mugnolo \& Colley (2006) which minimize distance errors.

No map projection of the sphere can be perfect.  But projections can
conserve some properties.  The Mercator Projection is conformal
(preserving shapes locally), while the Mollweide projection (an
elliptical projection used by the WMAP team) preserves areas.

For conformal projections, Chebyshev (1856) showed the rms local scale
errors over a map are minimized when the scale factor on the boundary
of the region is a constant (e.g., Snyder 1993).  Thus the conformal
map of a hemisphere with the smallest rms logarithmic scale errors is
the Stereographic projection.  The Hammond atlas (1992) has developed
optimal conformal projections for individual continents using the
Chebyshev criterion of constant scale factor on the boundary.  Such a
conformal map of the sphere with a boundary cut from pole to pole
along the $180^\circ$ meridian of longitude is the Eisenlohr
projection where the scale factor at the boundary is ($3 + 2\sqrt{2}$)
times larger than at the center (c.f Snyder 1993).  But this is not
necessarily the optimal solution. An eastern and western hemisphere
side by side with Stereographic projections would produce a conformal
map with a scale factor that varied from the center to the
circumference of each hemisphere by a factor of only $\sqrt{2}$.
Indeed, by making map projections with more and more interruptions
(like the Goode interrupted projection) one can make the scale errors
as small as one pleases.

But map projections with many interruptions are unappealing.  Why?
Boundary cuts can intersect geodesics connecting random points on the
globe (Goldberg \& Gott 2006).  Also, importantly, interruptions take
points that are close together on the sphere and put them far apart on
the map.  The distances between those pairs of points are quite
inaccurate. Maps show distances between points and so one also might
want to minimize the errors in these distances.  This is a global
rather than a local criterion and appropriately penalizes projections
with too many interruptions.  Every map should have a scale bar at the
bottom. If two points are shown twice as close on the map as they are
on the globe that is just as bad an error as if they were shown twice
as far away, so it is the rms logarithmic distance errors between
random points on the sphere that we wish to minimize.

Gott, Mugnolo \& Colley (2006) have measured the distance errors in
different map projections by placing 30,000,000 random pairs of points
on the globe and measuring the rms logarithmic distance error between
these pairs of points on the map versus the distance between these
pairs of points on the globe.  (In each case, the overall size of the
map relative to the map scale is adjusted to minimize the rms scale
error first.) The Mercator projection they find has $\sigma = 0.444$.
Thus we may simply say that the Mercator projection has distance
errors of 44.4\% where the phrase ``the rms logarithmic distance error
= 0.444 = 44.4\% between pairs of random points'' is understood.  By
comparison, the Mollweide projection is better, with distance errors
of 39.0\%. It avoids the overly large polar areas of the Mercator and
plots the north and south poles as points.  The Winkel-Tripel
projection currently used by the National Geographic Society for its
world maps has distance errors of 41.2\%.  The Hammer equal-area
projection has errors of 38.8\%.

\begin{figure}
\caption[]{WMAP three-year data (ILC map) for the celestial sphere using the
Gott equal-area elliptical projection with distance errors of 36.5\%.
The Gott-Colley (2003) color scheme is used.  The average temperature
is plotted as white.  Above average temperatures are plotted in red,
with the amount of red ink proportional to the temperature difference
from the mean.  Below average temperatures are shown in blue, with the
amount of blue ink proportional to the temperature difference from the
mean.  This accurately portrays the symmetry between the hot and cold
spots.  The galactic center is in the center of the map, the galactic
plane is a horizontal line-the major axis of the ellipse.  The north
galactic pole is at the top of the map, and the south galactic pole is
at the bottom.  Shapes are shown properly along the central meridian.
Since the CMB fluctuations are isotropic, one can see the shape
distortions easily as one goes away from the central meridian and
toward the edges of the map.}
\label{WMAP_gottell}
\end{figure}

The new Gott equal-area elliptical projection has distance errors of
36.5\% as compared with the Mollweide equal-area projection which has
distance errors of 39.0\%.  The WMAP data are shown in this new
projection in Fig.~\ref{WMAP_gottell}.  The Mollweide projection has
perfect shapes locally at only two points on the central meridian of
the map, while the Gott elliptical projection has perfect shapes
locally along the entire central meridian.  For comparison, the Hammer
equal-area projection (with distance errors of 38.8\%) has perfect
shapes locally only at one point in the center of the map, while the
Eckert VI equal-area projection has distance errors of 38.5\%, and the
sinusoidal equal-area projection has distance errors of 40.7\%.  Thus,
the new Gott equal-area elliptical has smaller distance errors than
these other standard projections and has some nice properties in
addition.  Azimuthal projections offer the possibility of being best
overall.  The equidistant azimuthal projection has distance errors of
35.6\%, and the Lambert equal-area azimuthal projection has distance
errors of only 34.3\% (see Fig.~\ref{WMAP_lagrange}).

The new Gott-Mugnolo azimuthal projection has still smaller distance
errors of 34.1\%, which is the smallest of all map projections we have
studied.  Its radial distribution is designed to produce minimal
distance errors (see the appendix).  We present the WMAP data in this
projection in Fig.~\ref{WMAP_gottaz}.

\begin{figure}
\caption[]{WMAP three-year ILC data, using the Lambert equal-area
  azimuthal projection, with distance errors of 34.3\%.  This gives a
  view of the CMB as reflected in a spherical mirrored garden ball
  seen from a great distance.}
\label{WMAP_lagrange}
\end{figure}

\begin{figure}
\caption[]{WMAP three year data, using the Gott-Mugnolo azimuthal
projection with distance errors of 34.1\%.  We have placed the south
galactic pole at the center of the map. The north galactic pole is at
the circumference of the map.  The galactic equator is a circle
centered on the south galactic pole whose radius is 65.4\% of the
radius of the map.}
\label{WMAP_gottaz}
\end{figure}

\section{Genus of the WMAP 3-year Data}

The most direct data product from the WMAP team is the ``internal
linear combination'' (ILC) map of the CMB, which is given in the
HEALPix format (G\'orski~\etal\ 2000).  This ILC map uses the optimum
linear combination of the skymaps at the different frequencies to
remove the Galaxy and some other foregrounds (LAMBDA ILC 2006).  Note
that the method is significantly different than the map used in the
1-year data release (Bennett \etal\ 2003b), though details of the
difference are not perfectly clear.  This is the primary map
distributed as the best rendering of the CMB and has a resolution
(beam width) of $1^\circ \times 1^\circ$.  The WMAP team has also
produced (as with the 1-year data) a set of masks that can be used to
exclude pixels regarded as contaminated by a foreground, Galactic or
otherwise.  We have chosen the {\it Kp0} mask provided by the WMAP
team.

Colley \etal\ 2003 describe in detail the computation of the genus on
the sphere under a HEALPix map projection.  For 2D topology on a
plane, the 2D genus of the microwave background is defined as
\begin{equation}
g_{2D} = \mbox{number of hot spots} - \mbox{number of cold spots}.
\end{equation}
For a Gaussian random-phase field,
\begin{equation}
g_{2D} \propto \nu\exp(\nu^2/2),
\end{equation}
where $\nu$ measures the number of standard deviations above the mean
temperature (for $\nu > 0$, there are more hot spots than cold spots,
and for $\nu < 0$, there are more cold spots than hot spots).  The
genus is also equal to the integral of the curvature around the
temperature contour divided by $2\pi$.  If we were to drive a truck
around an isolated hot spot, we would have to turn a total angle of
$2\pi$ as we completed a circuit around the hot spot.  Driving a truck
around an isolated cold spot, we would turn a total angle of $2\pi$ in
the opposite sign, with a negative turn angle defined as one that is a
turn to the left when the hot region is on your right.  This has been
carried out on planar images by the program CONTOUR2D which counts
the turning observed at each vertex of four pixels in an image
(Melott~\etal\ 1989).

We can rigorously define the 2D genus on a spherical surface
(Colley~\etal\ 2003).  The 2D genus is defined to be equal to minus
the 3D genus of solid objects formed by bestowing the hot spots with a
small, but finite radial extent.  Imagine using lead paint to paint
the hot regions onto the surface of a balloon, and after letting the
paint dry, bursting the balloon to obtain solid, curved lead shapes
that would have a certain 3D genus.  Take the minus of this number and
that will be $g_{2D}$, as we will define it.

One hot spot in the north polar region would have a 2D genus of +1
(one hot spot), because the hot spot cap is one isolated region.
Suppose the hot region covered all of the sphere except for a cold
spot in the south polar region.  The genus would still be +1, because
this would look like a sugar bowl without any handles, which is also
one isolated region in 3D.  The topology in each case is identical
since one can be deformed into the other.  The genus on a plane is
determined by the turning that a truck would do driving around the
temperature contour surface.  Circling a hot spot on a plane would
require a total turning of $2\pi$.  The Gaussian random-phase formula
measures this local turning.  Circling a hot spot on the sphere
involves a total turning of $2\pi - 4\pi f$, where $f$ is the fraction
of the sphere in the hot spot (because the deficit produced by
parallel transport on the sphere is equal to the enclosed area).
Dividing by $2\pi$, we may define the effective genus:
\begin{equation}
\gtwodeff = g_{2D} - 2f,
\label{eq_g2Deff}
\end{equation}
where $f$ is the fraction of the area of the sphere in the hot spots.
For a Gaussian random-phase field on the sphere
\begin{equation}
\gtwodeff \propto \nu\exp(-\nu^2/2),
\end{equation}
because the Gaussian random-phase field behaves locally on the sphere
just as it does on the plane to produce this particular contribution
to the turning integral.  Thus, in comparing the WMAP data to the
random-phase formula, we will use $\gtwodeff$, as defined rigrously
above.

We have implemented this method on the HEALPix map projection as
follows.  Within any of the 12 principal diamonds on the projection,
the method is as straightfoward as on any planar image.  However, much
care must be taken where diamonds meet so as not to overcount or
undercount any pixel vertices.  Also, in 8 places on the sphere, three
diamonds meet at a single vertex (such corners contain all of the
curvature in the projection); these special vertices are addressed
carefully in Colley~\etal\ 2003.  Masking a portion of the sphere
means the contours around various structures may be cut off randomly
by the mask.  However, this randomness is what we rely on.
Colley~\etal\ 2003 show that as long as the the mask does not
preferrentially excise higher or lower temperature pixels than
average, the genus will be, on average, unaffected, although the
number of structures will be expected to decline by a fraction equal
to the fraction of the total sphere excised by the mask.

\begin{figure}
\resizebox{\hsize}{!}{\includegraphics{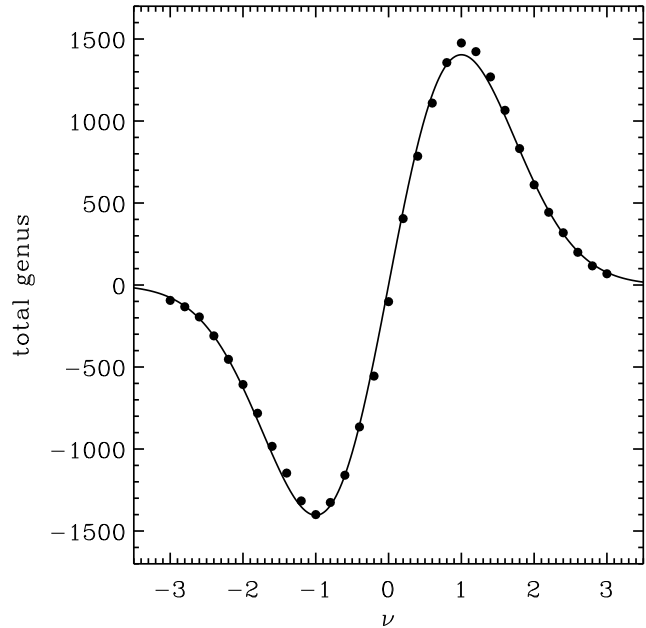}}
\caption[]{Total genus, $g_{2D\!,\eff}$, in the WMAP 3-yr ILC
  (internal linear combination), which is smoothed to a FWHM of 1
  degree.  HEALPix pixels excluded by the suggested {\it Kp0} mask
  have been omitted in the genus computation.  The solid curve shows
  the best-fit random-phase curve $g(\nu) \propto \nu\exp(-nu^2/2)$.}
\label{realgenfig}
\end{figure}

Fig.~\ref{realgenfig} shows the total genus of the masked WMAP 3-year 
data.  We do not provide errorbars (most are so small they would be
inside the plot points anyway), because, as we will soon show, simple
independent errorbars are not sufficient to convey the information on
the quality of fit to the curve.

Colley (1997) introduced a new method for computing the confidence of
fit for the 2-D genus curve.  In that work, the genus of the galaxy
distribution as observed in the Las Campanas Redshift Survey
(Schechtman~\etal\ 1996) was computed.  To characterize the quality of
fit of the LCRS genus to the 2-D theoretical genus curve, the author
created 100 fake 2-D random-phase fields with a a power-spectrum of $n
= -1$, then cut out of those maps fan shapes that matched the fanned
shapes of the 6 redshift slices in the LCRS.  With these 100 fake
density maps, one could perform genus computations on each, and, as
with the real data, find the best-fit Gaussian random-phase genus
curve.  The errors with respect to that curve, at each value of $\nu$
can be cross-correlated to give a covariance matrix $C(\nu_1,\nu_2)$
which contains the expected product of the errors in the genus at
$\nu_1$ and $\nu_2$.  Very similar analysis was carried out for the
IRAS 1.2-Jansky Redshift Survey by Protogeros \& Weinberg (1997)
nearly coincidentally with Colley's work.

Spergel~\etal\ 2006 carry out a very similar calculation in their
analysis of the genus.  One very important distinction is that those
authors compare their genus values directly to the means measured in
their fake maps.  Based on the best-fit flat-lambda model, they find
that the WMAP data produce a genus curve that looks excellent but has
a high amplitude compared to the mean genus of the fake maps.  The
amplitude of the genus curve depends only on the shape of the power
spectrum, so the errors they found are primarily due to errors in the
power spectrum (if nothing else, the quadrupole is low relative to
that predicted by the best-fit flat-lambda model, predicting a genus
amplitude that is too high), rather than errors in the random-phase
nature of the temperature distribution.  Because the amplitude is a
function only of the power-spectrum, which we are not interested in
here, we do not impose an amplitude constraint.

We carry out an alternate analysis herein.  First, we created 200 maps
of fake CMB signal with the same power spectrum as that measured by
the WMAP team for the 3-year data release (Spergel~\etal\ 2006).  We
computed the genus, just as we did in Fig.~\ref{realgenfig} for all
200 maps.  In this case, we simply took the mean genus at the 31 $\nu$
values from all 200 maps and evaluated the errors in each map with
respect to that mean.  Again, multiplying the error at $\nu_1$ by the
error at $\nu_2$ gives the covariance $C(\nu_1,\nu_2)_j$ for each map
$j$.  Averaging this matrix over the 200 maps gives us an excellent
estimate of the covariance matrix, $C(\nu_1,\nu_2)$ with which to
carry out confidence of fit computations.

\begin{figure}
\resizebox{\hsize}{!}{\includegraphics{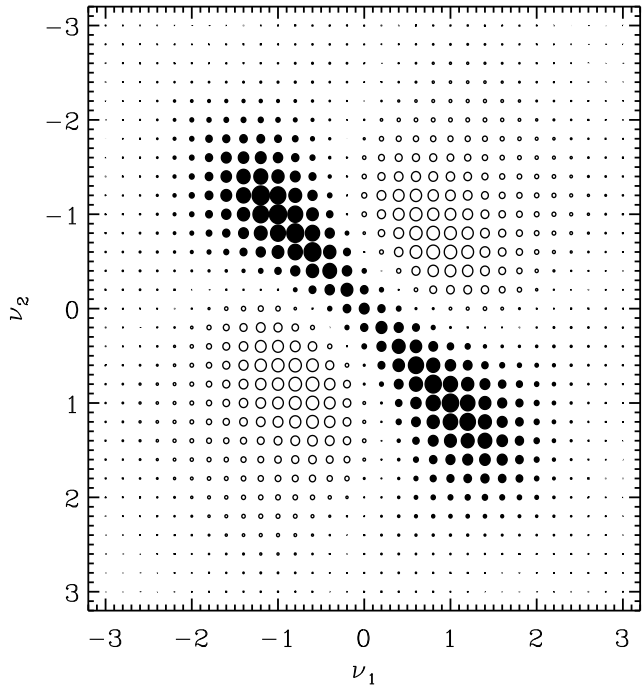}}
\caption[]{Covariance matrix of the genus measured at different values
  of $\nu$, as derived from the 200
  simulated CMB datasets.  Covariance is proportional to symbol size,
  and negative covariances are shown as open symbols. }
\label{covfig}
\end{figure}

As did Spergel~\etal\ (2006), we observe that the genus amplitude in
our fake maps is not as high as in the real WMAP data.  Since the
amplitude is a function only of the power-spectrum, there are two
possible sources of this discrepancy: mis-estimation of the
power-spectrum, and noise in the real data.  We are only using the
fake maps to produce a reasonable characterization of the errors
(covariance matrix), so we need only concern ourselves with how well
genus deviations in the fake maps correspond to the expected
deviations in the real data.  To investigate this issue, we have
carried out a great many Monte Carlo experiments using synthetic
Gaussian random-phase maps with various power-spectra to estimate the
effect on the covariance matrix when small changes in the
power-spectrum and pixel noise are introduced.  To very good
approximation, the only effect is to multiply the original covariance
matrix by a factor equal to ratio of the new amplitude to the original
amplitude.  Specifically, the greatest change to any element of the
covariance matrix, other than this ratio, is typically of the order of
the fractional difference of the ratio and 1---if the amplitude
increases by 10\%, then the worst discrepancy of a single covariance
matrix element from the original covariance multiplied by 1.1, is of
order 10\%.  Most of the elements, in fact, do much better.  We
therefore regard our best estimate of the covariance matrix for the
WMAP data as the covariance determined from the 200 fake CMB maps,
multiplied by the ratio of the amplitude of the real data vs. the mean
amplitude from the fake maps.  The resulting covariance matrix is
illustrated in Fig.~\ref{covfig} (following Protogeros \& Weinberg
1997).  The size of the circles is proportional to the covariance;
filled circles indicate positive covariance, and open circles indicate
negative covariance.

\begin{figure}
\resizebox{\hsize}{!}{\includegraphics{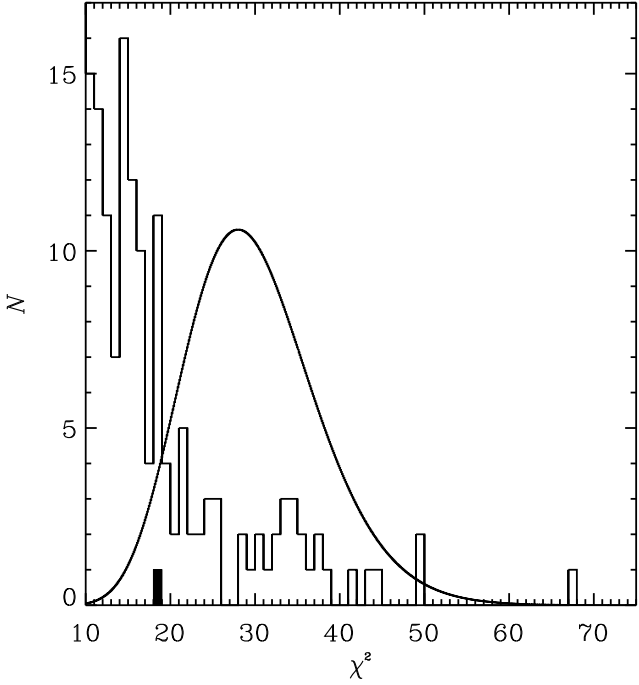}}
\caption[]{ Histogram of $\chi^2$ values for genus derived from the
  200 simulated CMB datasets (line), and for the genus derived from
  the real WMAP data (solid box).  $\chi^2$ values are
  computed assuming a diagonal-only covariance matrix (i.e., errors
  are regarded as independent among the $\nu$-values).}
\label{chi2diagfig}
\end{figure}

Before using this covariance matrix to carry out goodness of fit
calculations, we shall first examine how necessary it is to go to such
lengths.  We can force the covariance matrix to be diagonal by
disregarding the off-diagonal terms, and inserting along the diagonal,
the squares of the direct one-sigma errorbars observed at each genus
point among the 200 fake maps.  Computing the best fit amplitude for
each map, and the corresponding $\chi^2$ in the ``usual'' way used for
independent datapoints yields Fig.~\ref{chi2diagfig}.  The histogram
of the $\chi^2$ values are shown in outline.  Overplotted is the
expected distribution of 200 $\chi^2$ values for 30 degrees of freedom
(31 minus the amplitude fit).  Obviously, the performance is terrible
(because the individual data points are not independent).  Plotted as
a solid box is the WMAP data (whose errorbars have been scaled by the
square-root of the amplitude ratio, as discussed above).  The WMAP
genus, according to this ostensibly inept statistic, acquires a
naively good value of $\chi^2 = 18.0$, but this is actually just so-so
with respect to the $\chi^2$ values from the fake maps, coming in at a
rank of 142nd out of the 201 total maps (200 fake maps + 1 real map).

\begin{figure}
\resizebox{\hsize}{!}{\includegraphics{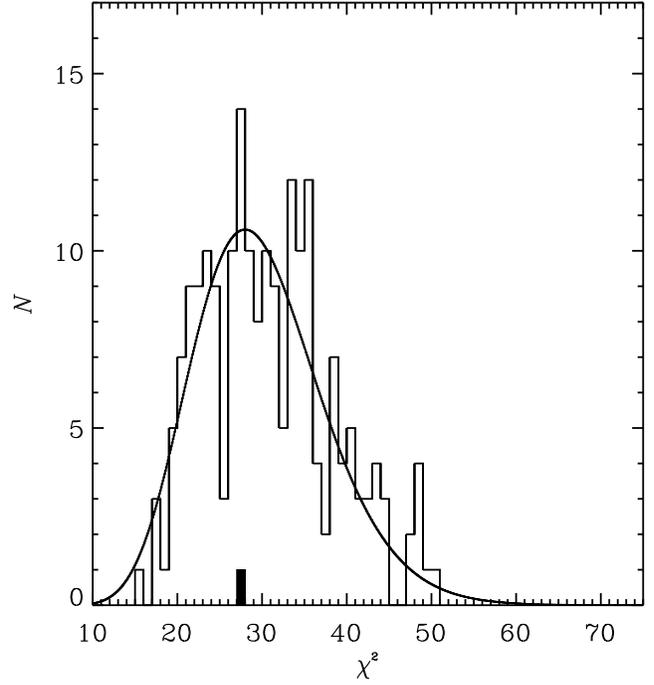}}
\caption[]{ As Fig.~\ref{chi2diagfig}, except that the full covariance
  matrix (see Fig.~\ref{covfig}) is used to compute the values of $\chi^2$.}
\label{chi2fig}
\end{figure}

After that aside, we carry on with the full covariance treatment of
the WMAP genus.  First, we find the best fit amplitude by minimzation
of $\chi^2$ as usual.  For a full covariance matrix, $C$, the best fit
amplitude is given as
\begin{equation}
A_{\mbox{\scriptsize best}} = { {\vec{g}\cdot C^{-1} \vec{g}_{2D}}\over
  {\vec{g}_{2D}\cdot C^{-1} \vec{g}_{2D}} },
\end{equation}
where $\vec{g}$ is the vector of measured genus values over the
selected $\nu$ values, and $g_{2D,i} = \nu_i\exp(-\nu_i^2/2)$, the
analytic form of the Gaussian random-phase two-dimensional genus curve
for the same $\nu_i$ values---note that this formula would be slightly
more complicated were $C$ not symmetric.  $\chi^2$ is computed in full
covariance matrix form as follows,
\begin{equation}
\chi^2 = \left[\vec{g}-A_{\mbox{\scriptsize best}}\vec{g}_{2D}\right] \cdot C^{-1} 
\left[\vec{g}-A_{\mbox{\scriptsize best}}\vec{g}_{2D}\right].
\end{equation}
Fig.~\ref{chi2fig}, as with Fig.~\ref{chi2diagfig} shows the histogram
of $\chi^2$ values for the 200 fake maps in outline, and the expected
distribution of 200 $\chi^2$ variates with 30 degrees of freedom as
the solid curve.  Notice that the $\chi^2$ values fit the expected
distribution much, much better than in Fig.~\ref{chi2diagfig}.  The
solid block in Fig.~\ref{chi2fig} gives the locus of the WMAP $\chi^2$
value of 27.38.  The cumulative probability that a 30 d.o.f. $\chi^2$
variate would exceed this value is 60\%, while the rank of the WMAP
data among the 201 total maps is 73rd, both excellent confirmations of
the Gaussian random-phase hypothesis.

\section{Non-linearity}

Various theories for non-linear perturbations in the CMB temperature
maps can be approximated as a simple 2nd-order correction to the linear
theory (Spergel et al. 2006).  On the scales we are considering (i.e.,
a smoothing length of $1^\circ$), the Sache-Wolfe (1967) effect is
dominant, so
\begin{equation}
{{\Delta T}\over{T}} = -{1\over 3}\Phi,
\end{equation}
where $\Phi$ is the curvature perturbation.  If we were considering
smaller smoothing scales, velocity effects would have to be considered
and on much larger scales, the integrated Sachs-Wolfe effect would
have to be considered.  Following Spergel~\etal\ (2006),
\begin{equation}
{{\Delta T}\over{T}} = -{1\over 3}\Phi(\vec{x}) = -{1\over 3}\left[\psi(\vec{x}) + f_{nl}\psi^2(\vec{x})\right],
\end{equation}
where $\psi(\vec{x})$ is a Gaussian random-phase field that is simply
a scaled version of of the unit Gaussian random-phase field
$\nu(\vec{x})$:  $\psi(\vec{x}) = -\sigma_\psi\nu(\vec{x})$, and
$f_{nl}$ is the amplitude of the non-linear effects.
The temperature map in the CMB is thus given by
\begin{equation}
  {{\Delta T}\over{T}} = -{1\over 3}\Phi = 
  {1\over 3}(\sigma_\psi\nu - f_{nl}\sigma_\psi^2\nu^2).
\end{equation}

Of course, the quadratic term will shift the observed mean and change the
standard deviaton.
\begin{equation}
\begin{array}{rl}
  \left\langle{{\Delta T}\over{T}}\right\rangle = &
	    -{1\over 3}f_{nl}\sigma_\psi^2\\
	    \left\langle{\left({\Delta T}\over{T}\right)^2}\right\rangle = &
	    {1\over 9}\sigma_\psi^2 +
	    {1\over 3}f_{nl}^2\sigma_\psi^4,\\
\end{array}
\end{equation}
the second of which can be used to compute the true standard deviation
in $\psi$ ( = $\sigma_\psi$) from the observed standard deviation in
tempature and $f_{nl}$.

When computing the genus in the 
usual way (subtract the mean and normalize by standard deviation), one
would observe
\begin{equation}
  \nu_{obs} = { { { {{\Delta T}\over{T}} - \left\langle{{\Delta
     T}\over{T}}\right\rangle} }\over {
    \left\langle{\left({\Delta T}\over{T}\right)^2}\right\rangle^{1/2} }} =
     { { {1\over 3}\left(\sigma_\psi\nu - f_{nl}\sigma_\psi^2\nu^2\right) + 
     { {1\over 3}f_{nl}\sigma_\psi^2}
}\over{
       \sqrt{	    {1\over 9}\sigma_\psi^2 +
	    {1\over 3}f_{nl}^2\sigma_\psi^4}} }
\end{equation}
Inversion of this equation allows us to compute the true value of
$\nu$ from an observed $\nu_{obs}$: $\nu =
\nu(\nu_{obs},f_{nl},\sigma_\psi)$, which can be used in the usual 2D
random-phase genus formula, $g(\nu) \propto \nu\exp(-\nu^2/2)$.

\begin{figure}
\resizebox{\hsize}{!}{\includegraphics{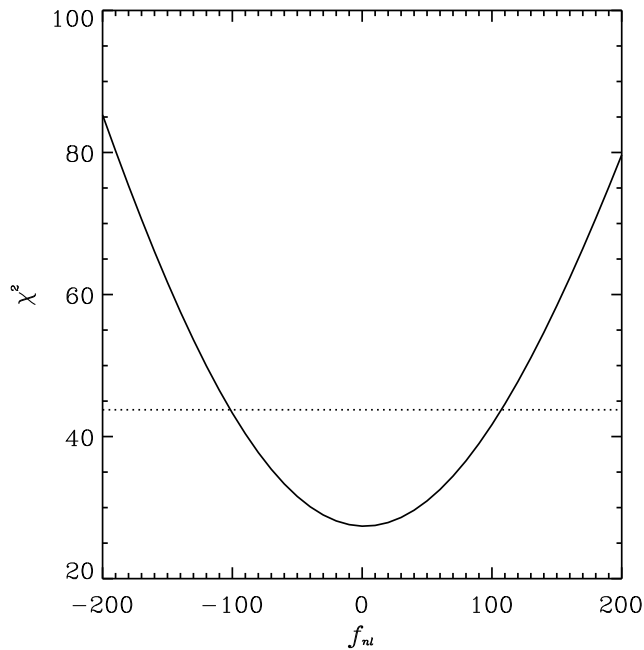}}
\caption[]{ $\chi^2$ as a function of $f_{nl}$.  The horizontal line 
  indicates the 95\% confidence interval for a $\chi^2$ variable with
  30 degrees of freedom.}
\label{fnl_chi2}
\end{figure}

The form of the analytic genus-curve, of course, changes under this
transformation, and so one can test the measured genus against the
non-linear genus curve by again using the $\chi^2$ test.
Fig.~\ref{fnl_chi2} shows the value of $\chi^2$ as a function of
$f_{nl}$.  Overplotted is the horizontal line at which the value of
$\chi^2$ is excluded at 95\% confidence for 30 degrees of freedom.  We
find the $f_{nl}$ is bounded to range between $-101$ and $+107$ to
maintain consistency with the observed genus curve at the 95\%
confidence level.  Notice that the minimum of the $\chi^2$ curve
occurs very close to $f_{nl} = 0$ which corresponds to the true
random-phase distribution with no non-linear effects.

\begin{figure}
\resizebox{\hsize}{!}{\includegraphics{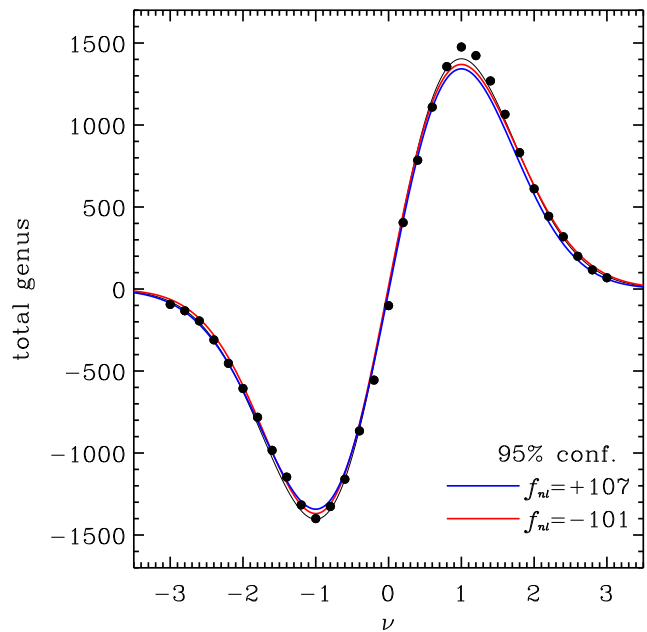}}
\caption[]{The genus curves for $f_{nl} = 0$ in black, for $f_{nl} =
  -101$ in red, and $f_{nl} = +107$ in blue.  The latter two of these
  represent the boundaries of 95\% confidence interval for $f_{nl}$ as
  shown in Fig.~\ref{fnl_chi2}.} 
\label{fnl_genus}
\end{figure}

Fig.~\ref{fnl_genus} shows how discrepant the non-linear genus curves
are from the observed genus values.  In red, we have $f_{nl} = -101$,
and in blue $f_{nl} = +107$.  In both cases, the best-fit amplitude
has decreased compared to the best-fit for the linear genus curve.
This is not surprising---when a less appropriate function is fit in
amplitude to a set of points, the amplitude is usually reduced.

Most of the power of this $\chi^2$ test for non-linearity lies not in
the peaks, where the amplitude change is most noticeable, but in the
wings of the genus curve.  Two factors are at work.  First, the
covariance matrix has very small entries for high and low $\nu$ (see
Fig.~\ref{covfig}), which makes any departure from the normal genus
curve more difficult there.  Second, most of the non-linear ``action''
is in the wings.  These two factors introduce sufficient statistical
power to allow the apparently slight departures from the Gaussian
random-phase curve in Fig.~\ref{fnl_genus} to be rejected with 95\%
confidence.

Our results are similar to those obtained for the WMAP one-year data,
$-54 \le f_{nl} \le 134$ (Komatsu~\etal\ 2003 in Spergel~\etal\ 2006)
at the 95\% confidence level.  Both support standard inflation which
predicts $f_{nl}$ of order unity due primarily to non-linear
gravitational effects at recombination; the contribution from
slow-roll inflation is much less, of order 0.01 (e.g., Komatsu \&
Spergel 2000; Maldacena 2005 and references therein). 

\section{Low $\ell$ modes}
The low $\ell$ power-spectrum of the CMB produced by WMAP has
generated substantial interest due to some fairly suprising behaviors.
The quadrupole moment appears to be strangely low, we have noticed
that the next several moments appear to oscillate in sawtooth fashion
about the best-fit flat-lambda power-spectrum.  This behavior reminds
one of ``ringing'' in Fourier space of a real feature in the sky with
a negative quadrupole moment.  Of course, the Galaxy produces a large
positive quadrupole, which must be very carefully subtracted off to
reveal the CMB behind it.  Could it be that the Galaxy has been
over-subtracted, causing a small quadrupole (and associated ringing)
in the low $\ell$ modes of the power-spectrum? (e.g., Park~\etal\
2006)

First, we simply look at the modes themselves, in real-space on the
sky.  Fig.~\ref{mergel8} (top) shows the first 8 modes of the
foreground given by the WMAP team.  The galaxy is obvious, as is
ringing at a galactic latitude of roughly $45^\circ$.  Next, we look
at the first 8 modes of a foreground cleaned map (Park \etal\ 2006) in
Fig.~\ref{mergel8} (bottom).  On first blush, the two maps show little
correspondence; however, a closer inspection reveals some apparent
anti-correlations.  Notice that the largest cold (blue) spot in the
cleaned map lies on the galactic plane, while the largest hot (red)
spot in the cleaned map (on the left) lies on the coldest point of the
galactic plane.  Also, there is some possible mischief at the $\pm
45^\circ$ bands in the cleaned map, which corresponds to the ringing
in the foreground map.

\begin{figure}
\resizebox{\hsize}{!}{\includegraphics{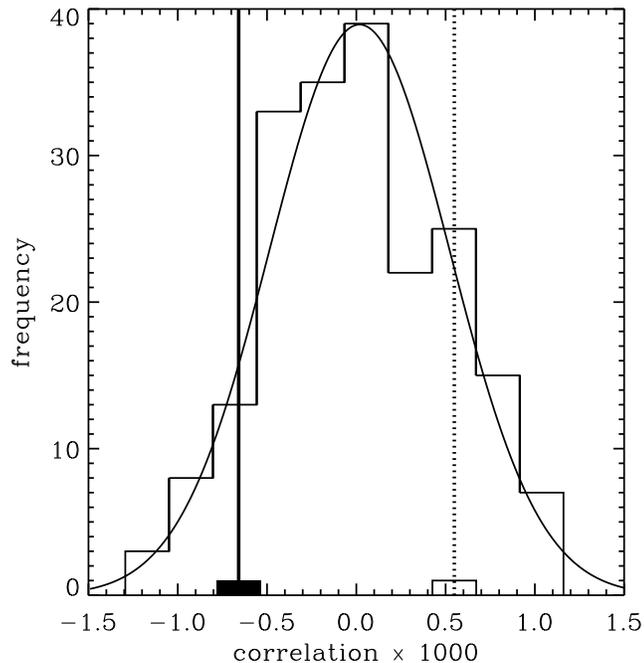}}
\caption[]{Correlations of a set of 200 fake $\ell \le 8$ maps with the
  WMAP foreground $\ell \le 8$ map (line histogram).  Shown as the
  black box and heavy vertical line on the left is the locus of the
  correlation of the cleaned ILC map at 
  $\ell \le 8$.  This reveals some anti-correlation.  The open box and dotted
  vertical line on the right show the locus of the correlation of the
  same map, rotated $90^\circ$.}  
\label{forecorr}
\end{figure}

To test whether these coincidences would have occurred randomly, we
conducted a correlation analysis.  First, we constructed 200 fake maps
with $\ell \le 8$, and correlated these with the foreground map, by
simply adding up the product of the pixel values from each map,
\begin{equation}
R = \sum_i^{\mbox{\scriptsize{all pix}}}
{\Delta
  T_i(\mbox{foreground})\cdot\Delta T_i(\mbox{map})}.
\end{equation}
Fig.~\ref{forecorr} shows as a solid histogram the correlation sums
for the 200 fake maps correlated with the real foreground.
Overplotted is a Gaussian with the standard deviation and mean of the
distribution of the correlations.  The same figure shows as a solid
block and heavy vertical line at $-0.65$ on the graph, the
anti-correlation of the cleaned map.  This anti-correlation is rare at
the $-1.35$-$\sigma$ level ($P = 8.9\%$), i.e., 91.1\% of random maps
would be less anti-correlated.  The value ranks 17 out of the 201
total datasets analyzed, so 184 out of 201 are less anti-correlated.
While this is a confirmation of what we see with our eyes, the value
falls well short of the two-sided 2-$\sigma$ threshold ($P =
2.275\%$), so we do not regard this anti-correlation as statistically
significant.  We further tested the correlation by rotating the
cleaned map by $90^\circ$, such that the origin of galactic
coordinates ($\ell{\mbox{\scriptsize II}} = 0^\circ,
b{\mbox{\scriptsize II}} = 0^\circ$) is placed at the north pole, and
re-ran the correlation.  Here, we find a positive correlation of 0.584
(shown on Fig.~\ref{forecorr} as an open box and dotted vertical
line).  This occurs at the 1.06-$\sigma$ interval ($P = 85.3\%$), and
is therefore also consistent with the distribution of fake map
correlations.  A separate, detailed analysis of foregrounds, masks and
correlations with the foregrounds in the low-$\ell$ modes has been
carried out by Oliveira-Costa \& Tegmark (2006).  These authors
confirm the notion that errors in the low-$\ell$ modes are largely due
to foregrounds.

\begin{figure}
\resizebox{\hsize}{!}{\includegraphics{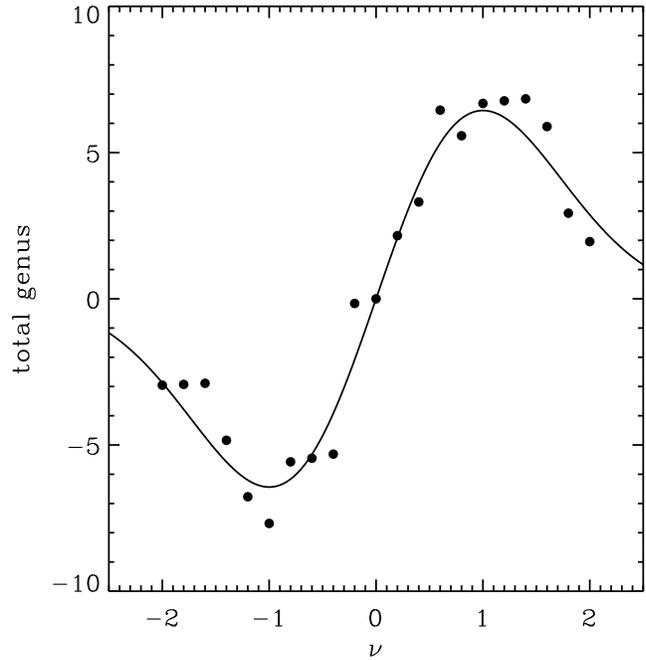}}
\caption[]{Genus, $\gtwodeff$, of the $\ell \le 8$ map, with the
  analytical random-phase genus curve after a best-fit for amplitude
  has been made.}
\label{L8gen}
\end{figure}

\begin{figure}
\resizebox{\hsize}{!}{\includegraphics{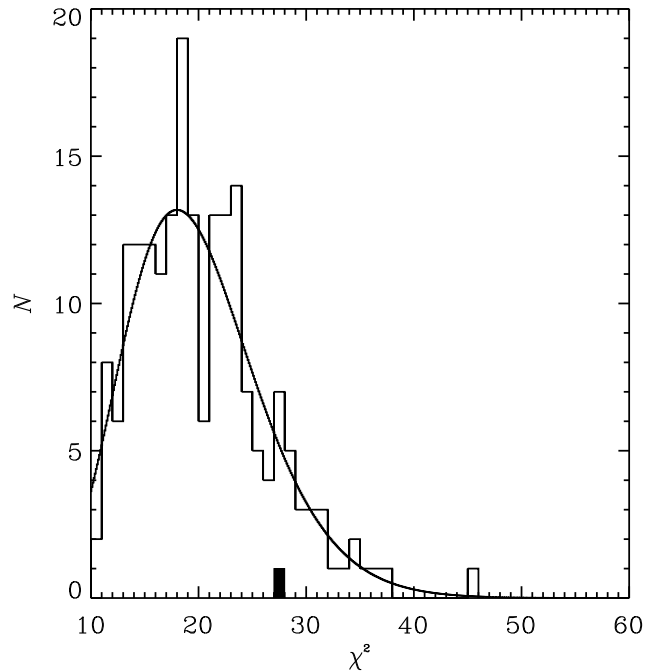}}
\caption[]{Histogram of $\chi^2$ values for fits of the analytical
  genus curve to the 200 fake $\ell \le 8$ maps (solid line) and for
  the cleaned ILC map (solid box).}  
\label{L8chi2}
\end{figure}

As a final test, we have measured the genus of the $\ell \le 8$
cleaned map.  Because there are few structures, we only test in the
range $-2<\nu<2$.  Whenever a low number of modes is encountered, as
in this case, it is essential to use Eq.~\ref{eq_g2Deff} rigorously to
account for the curvature of the sphere.  Fig.~\ref{L8gen} shows the
actual genus of the $\ell \le 8$ cleaned map as solid points, with the
best-fit analytic curve overplotted.  For statistical comparison, we
used the same 200 fake $\ell \le 8$ maps from above and measured their
genus values in the same range.  As we did with the 1-degree maps, we
constructed from these 200 maps the covariance matrix of the genus
values, so that we could measure $\chi^2$ formally for the real data.
Fig.~\ref{L8chi2} shows the histogram of the $\chi^2$ values for the
200 fake maps, and the overplotted $\chi^2$ distribution for 20
degrees of freedom.  The cleaned map produces a $\chi^2$ value of 27.8
(shown on the figure as a black box), which occurs at $P = 88.5\%$,
again well inside the 95\% (2-$\sigma$) limit.  The value ranks 178
out of the 201 datasets tested.  Both of these tests, therefore, show
that the low-$\ell$ modes observed by WMAP are consistent with the
Gaussian random-phase hypothesis.

\begin{figure}
\caption[]{Mollweide projection of the WMAP foreground (blue =
  $-0.4\mbox{mK}$, red =
  $+0.4\mbox{mK}$) {\it [top]}, and of the SILC cleaned map (blue =
  $-0.08\mbox{mK}$, red =
  $+0.08\mbox{mK}$) {\it [bottom]}.}  
\label{mergel8}
\end{figure}

\section*{Acknowledgments}
The WMAP datasets were developed by Princeton University and NASA's
Goddard Space Flight Center, and graciously provided to the public at
http://lambda.gsfc.nasa.gov/.  JRG thanks the National Science
Foundation for support under grant AST04-06713.  WNC would like to
thank NASA and the Center of Modeling, Simulation and Analysis for
providing computational services and indirect support for some of this
work.  CBP acknowledges the support of the Korea Science and
Engineering Foundation (KOSEF) through the Astrophysical Research
Center for the Structure and Evolution of the Cosmos (ARCSEC), and
through the grant R01-2004-000-10520-0.

\appendix
\section[]{New Map Projections}

The new Gott equal-area elliptical projection, designed by JRG to
lessen distance errors, is produced in the following way: Collapse the
longitudes by a factor of two toward the central meridian so the whole
globe is mapped onto one hemisphere of the globe.  Now establish east
and west ``poles'' $180^\circ$ apart on the equator, at the eastern and
western edges of this hemisphere, and define ``new longitude and
latitude'' relative to these two poles.  Then collapse the ``new
longitudes'' by a factor of two (toward the equator) so that the whole
globe is mapped onto a quadrant of the sphere.  The north pole is now
plotted at longitude $0^\circ$ and latitude $+45^\circ$, while the
south pole is plotted at longitude $0^\circ$ and latitude $-45^\circ$.
Both the first compression in longitude and the second compression in
``new longitude'' preserve relative areas, so the combination does as
well.  Then map this quadrant of the sphere onto a plane with a
transverse equal-area Bromley-Mollweide projection.  (The
Bromley-Mollweide projection is like the Mollweide
projection-elliptical, with elliptical longitude lines and straight
latitude lines, but stretched to produce an ellipse with an axis ratio
of $\pi^2/4:1$ so that the equator becomes a standard parallel where
shapes are preserved locally).  Since the quadrant being mapped is
bounded by two lines of ``new longitude'' and such lines are plotted as
ellipses by the transverse Bromley-Mollweide projection, the Gott
projection will map the earth onto an ellipse.  Since the longitude
and ``new longitude'' compressions are by a factor of two each in the
horizontal and vertical directions along the central meridian in the
map and the transverse Bromley-Mollweide projection preserves shapes
along this line, the Gott projection will preserve shapes locally
along the central meridian.  The map is an attractive ellipse with an
axis ratio of $16/\pi^2$:1 or 1.62211:1, close to the golden mean
(which is $[1 + \sqrt{5}]/2$:1 or 1.618).  The formulas for the Gott
equal-area elliptical projection are as follows. Cartesian map
coordinates $(x,y)$ may be calculated from the latitude and longitude
$(\phi,\lambda)$ (in radians) of a point on the globe by first defining
a ``new latitude'' and ``new longitude'':

\begin{equation}
\begin{array}{rl}
\phi^\prime =& -\arcsin\left[\cos\phi \cdot \sin(\lambda/2)\right]\\
\lambda^\prime =& 0.5 \arcsin\left(\sin \phi/\cos \phi^\prime\right)\\
\end{array}
\end{equation}

Then, 

\begin{equation}
\begin{array}{rl}
2\theta + \sin 2\theta =& \pi\sin\phi^\prime\\
x =& -\sqrt{2}\sin\theta\\
y =& {{\pi}\over{2\sqrt{2}}}\lambda^\prime\cos\theta\\
\end{array}
\end{equation}

\begin{figure}
\resizebox{\hsize}{!}{\includegraphics{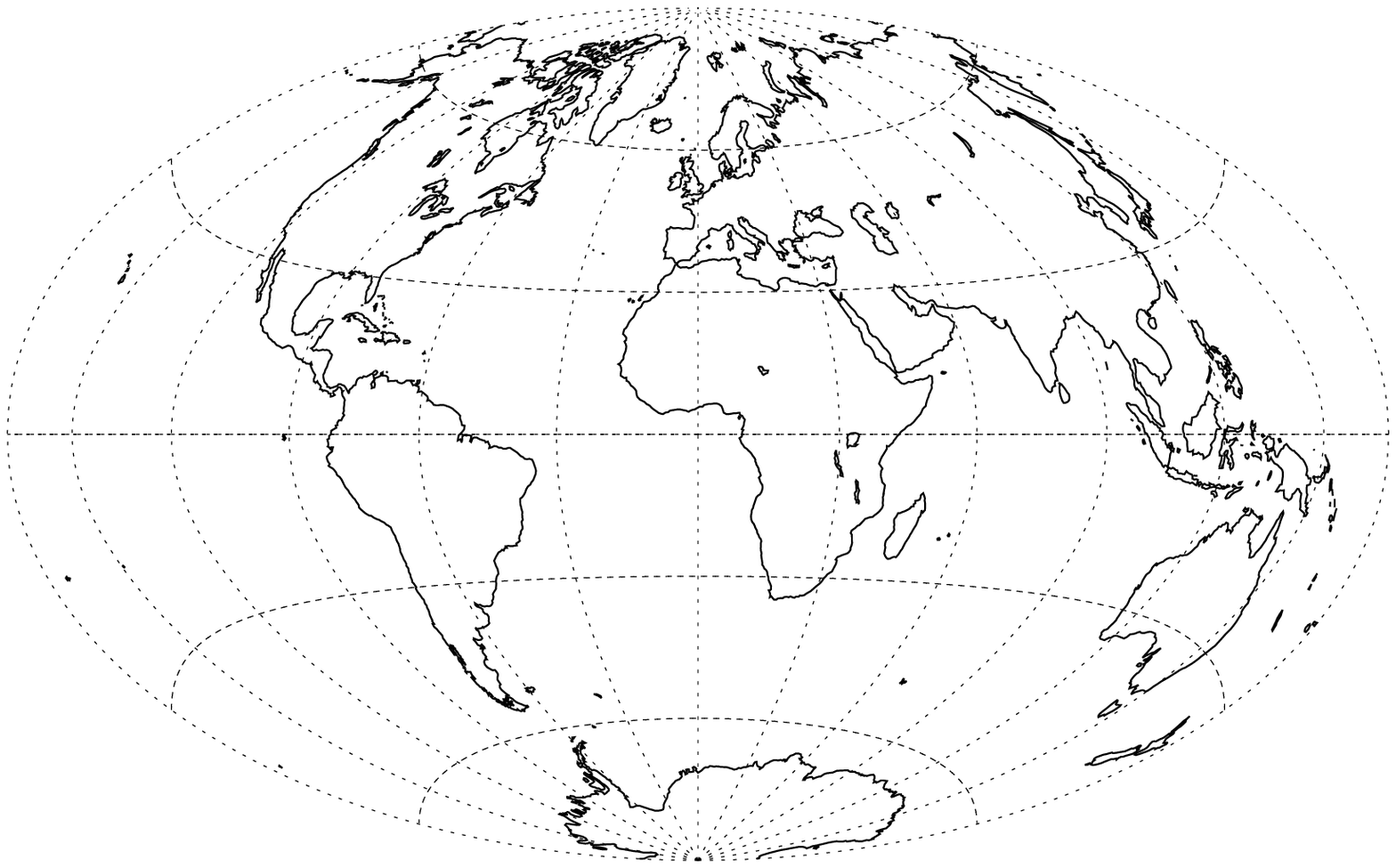}}
\caption[]{Map of the Earth using the Gott equal-area elliptical
projection.  Local shapes are perfect along the Greenwich Meridian}
\label{earth_gottell}
\end{figure}

In Fig.~\ref{earth_gottell} we have shown a map of the earth using
this new Gott equal-area elliptical projection.  Shapes are preserved
locally along the central meridian.  The distance scale is also linear
along this meridian.  This has good shapes for Europe, Africa, and
Antarctica.  The polar areas are better displayed than in the
Mollweide projection. Since the map is more nearly circular than the
Mollweide it makes smaller distance errors for points on opposite
sides of the international date line in the Pacific.  Also, the
lengths of the different meridians are more nearly equal on the map.
This has distance errors of 36.5\%.  For comparison,
Figs.~\ref{WMAP_gottell} and \ref{mars_gottell} show the Gott
equal-area elliptical projection of the WMAP 3-year data and the
planet Mars respectively.

\begin{figure}
\caption[]{Map of the Mars using the Gott equal-area elliptical
projection.  Local shapes are perfect along the central meridian.
Original imaging data from NASA/USGS (Viking Spacecraft).}
\label{mars_gottell}
\end{figure}

We were interested to see if the Lambert equal-area azimuthal
projection could be improved upon.  Thus Gott and Mugnolo placed 3,000
particles down randomly on the sphere and mapped them onto the plane
with the Lambert equal-area azimuthal projection.  They then
established a radial force between each pair of particles governed by
a potential which is proportional to the square of the logarithmic
distance error for that pair of particles on the map.  Each particle
is then allowed to move under the sum of the forces from the other
particles in the planar map.  This is an $N$-body problem in the
plane.  After one time step, their motion is stopped.  The total
potential energy of the system should be less than before.  This is
then repeated until the system settles into a relaxed distribution
where the potential and therefore the rms logarithmic distance errors
between pairs of points is in a local (and perhaps global) minimum.
The 3,000 particles give 4,498,500 pairs of distances.  The radial
distribution of points ($r$ as a function of latitude $\phi$) can then
be plotted.  This makes a tight scatter diagram which can be
approximated by the simple analytic formula:

\begin{equation}
r = \sin\left[0.446(\pi/2 - \phi)\right]
\end{equation}

(The formula for the Lambert equal-area azimuthal is similar except
0.446 is replaced by 1/2.)  Thus our formulae for the $(x,y)$
Cartesian coordinates on the Gott-Mugnolo azimuthal projection are:

\begin{equation}
\begin{array}{rl}
x =&\cos\lambda\sin\left[0.446(\pi/2 - \phi)\right] \\
y =&\sin\lambda\sin\left[0.446(\pi/2 - \phi)\right].\\
\end{array}
\end{equation}

Checking with 30,000,000 random pairs, we find that this map
projection has distance errors of only 34.1\% which is the lowest of
all the map projections we have studied.  The WMAP 3-year data are
shown in this map projection in Fig.~\ref{WMAP_gottaz}, (a
corresponding lunar map using this projection is shown in
Fig.~\ref{moon_gottaz}.)  The value of the optimal constant 0.446 was
then checked to an accuracy of two significant figures by varying this
constant and minimizing the errors.  Given the symmetry of the
problem, it is perhaps not surprising that the best projection of the
sphere for distance errors gives a map that is circular in shape.
These techniques may find future applications for maps of the earth
and of particular regions, as well as for the mapping of irregularly
shaped asteroids.

\begin{figure}
\caption[]{Map of the moon using the Gott-Mugnolo azimuthal
projection.  This has the smallest distance errors (34.1\%) of any
projection studied.  Original imaging data come from the NASA's
Clementine Satellite; the near face is in the center with the far side
as an annulus surrounding it.  The north and south poles are visible
as places where the shadows are prominent, because the sun is always
at low elevation there.}
\label{moon_gottaz}
\end{figure}

\label{lastpage}

\end{document}